\documentclass[12pt]{article}
\usepackage{graphicx}
\def \b{{\cal B}}

\def \bea{\begin{eqnarray}}
\def \beq{\begin{equation}}
\def \eea{\end{eqnarray}}
\def \eeq{\end{equation}}
\def \ite{{\it et al.}}

\begin{document}
\begin{flushright}
EFI 06-22 \\
TECHNION-PH-2006-17 \\
hep-ph/0610227 \\
October 2006 \\
\end{flushright}
\centerline{\bf SUM RULE FOR RATE AND CP ASYMMETRY IN $B^+ \to K^+ \pi^0$}
\bigskip
\centerline{Michael Gronau$^a$ and Jonathan L. Rosner$^b$}
\medskip
\vskip3mm
\centerline{$^a$\it Physics Department, Technion -- Israel Institute of
Technology}
\centerline{\it 32000 Haifa, Israel}
\medskip
\centerline{$^b$\it Enrico Fermi Institute and Department of Physics,
University of Chicago}
\centerline{\it Chicago, Illinois 60637, USA}
\bigskip
\centerline{ABSTRACT}
\medskip
A sum rule relating the ratio 
$R_c = 2 \Gamma(B^+ \to K^+ \pi^0)/
\Gamma(B^+ \to K^0 \pi^+)$ and 
the CP asymmetry 
$A_{CP}(B^+ \to K^+\pi^0)$ is proved to first order in the ratio of tree 
to penguin amplitudes.  The sum rule
explains why it is possible to have $R_c$ consistent with 1 together with a 
small CP asymmetry in $B^+ \to K^+ \pi^0$. The measured ratio 
$A_{CP}(B^+\to K^+\pi^0)/A_{CP}(B^0\to K^+\pi^-)$ rules out a small strong 
phase difference between 
a color-suppressed and a color-favored tree amplitude contributing to 
$B^+\to K^+\pi^0$ as favored by QCD factorization.
\bigskip

\noindent
PACS Categories:

\bigskip

Partial decay rates and CP asymmetries in $B \to K \pi$ decays have been
mapped out with increasing accuracy thanks to high-statistics studies by the
BaBar \cite{Bona,DiMarco,Aubert:2006gm} and Belle \cite{Unno} Collaborations.
The experimental situation for branching ratios and CP asymmetries
is summarized in Table \ref{tab:bras}~\cite{HFAG,Aubert:2006ad,Abe:2006gy}, 
defining asymmetries by $A_{CP}\equiv [\Gamma(\bar B \to \bar f) 
- \Gamma(B \to f)]/[\Gamma(\bar B \to \bar f) + \Gamma(B \to f)]$.
These decays are dominated by an effective $\bar b \to \bar s$ penguin
amplitude, whose isospin-preserving property leads to the rate predictions
for $B$ and $\bar B$ decays,
\beq \label{eqn:rates}
\Gamma(B^+ \to K^0 \pi^+) = 2 \Gamma(B^+ \to K^+ \pi^0) = \Gamma(B^0 \to
K^+ \pi^-) = 2 \Gamma(B^0 \to K^0 \pi^0)~~.
\eeq

Tests for non-penguin amplitudes are provided by useful CP-averaged ratios
\cite{Fleischer:1997um,Buras:1998rb,Neubert:1998pt,Gronau:2001cj},
\bea
R & \equiv & \frac{\bar\Gamma(B^0 \to K^+ \pi^-)}
{\bar\Gamma(B^+ \to K^0 \pi^+)} = 0.92 \pm 0.07~~,\nonumber\\
R_c & \equiv & \frac{2\bar\Gamma(B^+ \to K^+ \pi^0)}
{\bar\Gamma(B^+ \to K^0 \pi^+)} = 1.11 \pm 0.07~~,\nonumber\\
R_n & \equiv & \frac{\bar\Gamma(B^0 \to K^+ \pi^-)}
{2\bar\Gamma(B^0 \to K^0 \pi^0)} = 0.99 \pm 0.07~~,
\eea
where $\bar \Gamma(B\to f)\equiv [\Gamma(B\to f)+\Gamma(\bar B\to \bar f)]/2$.
We have converted ratios of branching ratios to ratios of rates in the first
relation using the ratio $\tau_+/\tau_0 = 1.076 \pm 0.008$ of $B^+$ and
$B^0$ lifetimes \cite{HFAG}. At this time none of the three measured ratios 
provides a statistically significant evidence for a small $\Delta I \ne 0$
amplitude. 
Potental deviations from Eq.~(\ref{eqn:rates}), providing evidence for such an 
amplitude, are not expected to lead to a violation of a more general isospin sum 
rule obeyed by the four $B\to K\pi$ decay rates~\cite{Gronau:1998ep,Lipkin:1998ie}. 
This sum rule which  
is obeyed in the Standard Model up to second order corrections of non-penguin amplitudes,
holding also in the presence of first order isospin-breaking corrections~\cite{Gronau:2006eb}, 
is equivalent to the relation $R_c=R_n$~\cite{proof}.

Evidence for a non-penguin contribution and for a relative strong phase
between it and the penguin amplitude is provided by the non-zero direct CP
asymmetry $A_{CP}(B^0 \to K^+ \pi^-) = -0.093 \pm 0.015$.  On the other hand,
$A_{CP}(B^+ \to K^+ \pi^0)$ is consistent with zero, whereas one might have
expected it to be equal to $A_{CP}(B^0 \to K^+ \pi^-)$ if only the penguin
amplitude $P$ and a color-favored tree amplitude $T$ contributed to
both processes \cite{Gronau:1998ep}.  In fact, $B^+\to K^+\pi^0$ obtains also a
contribution from a color-suppressed tree amplitude $C$~\cite{Gronau:1994rj}.
There are now several arguments for $C$ to be large, comparable in magnitude to
$T$. The arguments include a global SU(3) fit to rates and asymmetries in $B\to
K\pi$ and $B\to\pi\pi$~\cite{Chiang:2004nm}, two separate analyses of
$B\to K\pi$~\cite{Baek:2004rp} and $B\to \pi\pi$~\cite{Buras:2003dj},
and recent calculations within perturbative QCD~\cite{Li:2005kt} and QCD 
factorization~\cite{Beneke:2005vv}. A significant amplitude $C$ contributing in 
$B^+ \to K^+ \pi^0$ may lead to a net small and even positive CP asymmetry in
this process if the interference terms of $T$ and $C$ with $P$ occur with
opposite signs~\cite{Lipkin:2006pc}.  Instead of equal asymmetries in $B^+\to
K^+\pi^0$ and $B^0\to K^+\pi^-$ one expects the leading terms in CP asymmetries
to obey an isospin sum rule relating all four $B \to K \pi$
processes~\cite{Atwood:1997iw,Gronau:2005kz}, or three decay modes, if one uses
the expectation that $A_{CP}(B^+ \to K^0 \pi^+)$ should be very
small~\cite{Gronau:2005gz}.

\renewcommand{\arraystretch}{1.4}
\begin{table*}
\caption{World averages of CP-averaged branching ratios $\b$ (in units of
$10^{-6}$) and direct CP asymmetries for $B \to K \pi$ decays taken from
Ref.~\cite{HFAG} unless quoted otherwise. 
\label{tab:bras}}
\begin{center}
\begin{tabular}{c c c} \hline \hline
Mode & $\b$ & $A_{CP}$ \\ \hline
$K^+ \pi^-$ & $19.7 \pm 0.6$ & $-0.093 \pm 0.015$ \\
$K^+ \pi^0$ & $12.8 \pm 0.6$ & $ 0.047 \pm 0.026$ \\
$K^0 \pi^+$ & $23.1 \pm 1.0$ & $ 0.009 \pm 0.025$ \\
$K^0 \pi^0$ & $10.0 \pm 0.6$ & $ -0.12 \pm  0.11$
\cite{Aubert:2006ad,Abe:2006gy} \\ \hline \hline
\end{tabular}
\end{center}
\end{table*}

A small asymmetry in $B^+\to K^+\pi^0$ implies bounds on the sine of the strong
phase difference $\delta_c$ between $T+C$ and $P$. The cosine of this phase
affects the ratio $R_c$ involving the decay rates for $B^+\to K^0\pi^+$ and
$B^+\to K^0\pi^+$. The question we wish to study in this note is whether the
two upper bounds on $|\sin\delta_c|$ and $|\cos\delta_c|$, from $A_{CP}(B^+\to
K^+\pi^0)$ and $R_c$, respectively, are consistent with each other. 
A potential inconsistency would be evidence for New Physics.
We will prove a sum rule involving both observables, in which an electroweak 
penguin amplitude plays an important role.  It will be shown that thanks to a
particular electroweak penguin contribution, the $B^+ \to K^+ \pi^0$ decay need
not display any evidence of non-penguin amplitudes through its rate {\it or}
its CP asymmetry. We will turn the argument around to update bounds on the weak
phase $\gamma\equiv {\rm arg}(-V^*_{ub}V_{ud}/V^*_{cb}V_{cd})$ using the
current measurements of $R_c$ and $A_{CP}(B^+\to K^+\pi^0)$.  We will also
study the ratio of asymmetries $A_{CP}(B^+\to K^+\pi^0)/A_{CP}(B^0 \to
K^+\pi^-)$ in terms of the strong phase difference between the amplitudes $C$
and $T$, showing that the measured ratio excludes a small phase difference
predicted by QCD factorization.

In order to prove the sum rule we write decay amplitudes in terms of
topological contributions~\cite{Gronau:1994rj},
\beq\label{eqn:amp}
A(B^+\to K^0\pi^+)  =  p + a~~,~~~~~-\sqrt{2}A(B^+\to K^+ \pi^0) 
=  p + t + c + a~~.
\eeq
Color-favored and color-suppressed electroweak penguin contributions,
$P_{EW}$ and $P^c_{EW}$, are included by defining~\cite{Gronau:1995hn} 
\beq\label{eqn:EWP}
p\equiv P-P^c_{EW}/3~~,~~~~~t \equiv T + P^c_{EW}~~,
~~~~c \equiv C + P_{EW}~~.
\eeq 
The annihilation amplitude $a$ can be safely neglected, as made evident by the
small CP asymmetry measured in $B^+\to K^0\pi^+$. (We are assuming that the
strong phase difference between $p$ and $a$ is not very small, as a sizable
$a$ would require rescattering.) 
 {\it A small $u$ quark contribution to the penguin
amplitude involving a CKM factor $V^*_{ub}V_{us}$ is absorbed in $T$ and $C$.}
In the standard phase convention~\cite{Yao:2006px} the amplitudes $P, P_{EW}$
and $P^c_{EW}$ involve a weak phase ${\rm arg}(V^*_{tb}V_{ts})=\pi$, while $T$
and $C$ carry a weak phase ${\rm arg}(V^*_{ub}V_{us})=\gamma$.

Flavor SU(3) symmetry relates electroweak penguin and tree amplitudes through a 
calculable ratio $\delta_{EW}$~\cite{Neubert:1998pt},
\bea\label{eqn:delta_EW}
t + c  & =  & T + C + P_{EW} + P^c_{EW} = (T + C)(1-\delta_{EW}e^{-i\gamma})~~,
\nonumber\\
\delta_{EW} & = & -\frac{3}{2}\frac{c_9 + c_{10}}{c_1 + c_2}
\frac{|V^*_{tb}V_{ts}|}{|V^*_{ub}V_{us}|} = 0.60 \pm 0.05~~.
\eea
The error in $\delta_{EW}$ is dominated by the current uncertainty in 
$|V_{ub}|/|V_{cb}| = 0.104 \pm 0.007$~\cite{Yao:2006px}, including also a 
smaller error from SU(3) breaking estimated using QCD factorization.
Eqs.(\ref{eqn:amp}) (\ref{eqn:EWP}) and (\ref{eqn:delta_EW}) imply
\cite{Gronau:2001cj}
\beq \label{eqn:Rc}
R_c = 1 - 2 r_c \cos \delta_c (\cos \gamma - \delta_{\rm EW})
+ r_c^2(1 - 2 \delta_{\rm EW} \cos \gamma + \delta_{\rm EW}^2)~,~~
\eeq
\beq \label{eqn:Acp}
A_{CP}(B^+ \to K^+ \pi^0) =  - 2 r_c \sin \delta_c \sin \gamma /R_c~~,
\eeq
where $r_c\equiv |T+C|/|p|$ and $\delta_c$ is the strong phase difference
between $T+C$ and $p$.  

The parameter $r_c$ is calculable in terms of measured
decay rates, using flavor SU(3) and noting that the tree amplitude $T+C$
dominates $B^+\to \pi^+\pi^0$ with a CKM factor $V^*_{ub}V_{ud}$ replacing
$V^*_{ub}V_{us}$ in $B^+\to K^+\pi^0$, up to negligible electroweak penguin
contributions.  Assuming factorization for this amplitude introduces a factor
$f_K/f_\pi$ for  SU(3) breaking, thus implying~\cite{Gronau:1994bn}
\beq
|T+C| = \sqrt{2}\frac{V_{us}}{V_{ud}}\frac{f_K}{f_\pi}|A(B^+\to \pi^+\pi^0)|~~.
\eeq
Using $\b(B^+\to \pi^+\pi^0) = (5.7 \pm 0.4)\times 10^{-6}$~\cite{HFAG} and 
taking $\b(B^+\to K^0\pi^+)$ from Table I, one finds
\beq 
r_c = \sqrt{2}\frac{V_{us}}{V_{ud}}\frac{f_K}{f_\pi}
\sqrt{\frac{\bar\b(B^+\to\pi^+\pi^0)}
{\bar\b(B^+\to K^0\pi^+)}} = 0.198 \pm 0.008~~.
\eeq
The error in $r_c$ does not include an uncertainty from assuming 
factorization for SU(3) breaking in $T+C$.
While this assumption should hold well for $T$, it may not be a good
approximation for $C$ which is more susceptible to final state interactions.
In fact, we will show below that the relative phase between $C$ and $T$ is not
small, contrary to a factorization prediction.  Thus one should allow a $10\%$
theoretical error when using factorization for relating $B \to K \pi$ and
$B \to \pi \pi$ $T+C$ amplitudes, so that
\beq
r_c =0.20 \pm 0.01~({\rm exp}) \pm 0.02~({\rm th})~~.
\eeq 

Eliminating $\delta_c$ in Eqs.~(\ref{eqn:Rc}) and (\ref{eqn:Acp}) by retaining
terms which are linear in $r_c$, one finds
\beq \label{eqn:sr}
\left( \frac{R_c-1}{\cos \gamma - \delta_{\rm EW}} \right)^2 +
\left( \frac{A_{CP}(B^+ \to K^+ \pi^0)}{\sin \gamma} \right)^2 = (2r_c)^2 +
{\cal O}(r_c^3)~~.
\eeq
The sum rule (\ref{eqn:sr}) implies that at least one of the two terms whose
squares occur on the left-hand-side must be sizable, of the order of
$2r_c=0.4$. The second term, $|A_{CP}(B^+\to K^+\pi ^0)|/\sin\gamma$, is
already smaller than 0.13 for $52^\circ \le \gamma \le 74^\circ$
\cite{CKMfitter} and using the current $2\sigma$ upper bound, 
$|A_{CP}(B^+\to K^+\pi^0)|<0.10$.
This bound implies a small value for $\delta_c, |\delta_c| < 20^\circ$.
The first term in (\ref{eqn:sr}) can 
saturate the sum rule with values of $R_c$ near 1 as
long as $\cos \gamma - \delta_{\rm EW}$ is small.  Since $52^\circ \le \gamma
\le 74^\circ$ implies $0.62 \ge \cos \gamma \ge 0.28$, this is easily arranged.
Thus, we conclude that thanks to the electroweak penguin contribution, the $B^+
\to K^+ \pi^0$ decay need not display any evidence of non-penguin amplitudes
through its rate {\it or} its CP asymmetry.

\begin{figure}
\begin{center}
\includegraphics[height=4.2in]{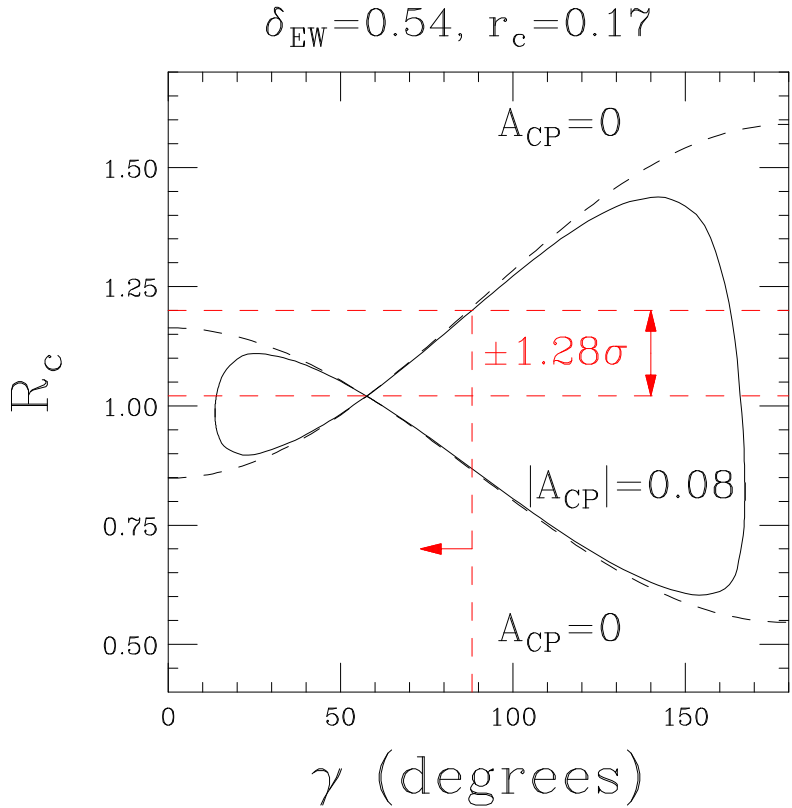}
\caption{Behavior of $R_c$ for
$A_{CP}(B^+\to K^+ \pi^0) = 0$ (dashed curves) or $|A_{CP}(B^+ \to K^+ \pi^0)|
= 0.08$ (solid curve) as a function of the weak phase $\gamma$.  Horizontal
dashed lines denote $\pm 1.28 \sigma$ experimental limits on $R_c$.
\label{fig:Rcacp}}
\end{center}
\end{figure}

In principle one could use Eq.\ (\ref{eqn:sr}) to place a bound on $\gamma$.
More precisely, the relations (\ref{eqn:Rc}) and (\ref{eqn:Acp}) correlate
$\gamma$, $R_c$, and $A_{CP}(B^+\to K^+\pi^0)$ \cite{Neubert:1998pt,
Gronau:2001cj}, as in the updated plot of Fig.\ \ref{fig:Rcacp}.  We show the
correlation that gives the {\it weakest} upper bound on $\gamma$, which
corresponds to the {\it lowest} values of $r_c$ and $\delta_{\rm EW}$.
Accordingly, we take the $-1.28 \sigma$ values $r_c=0.17$ and
$\delta_{\rm EW}=0.54$.  We show the cases $A_{CP}(B^+\to K^+\pi^0) = 0$ and
$|A_{CP}(B^+\to K^+\pi^0)| = 0.08$ which represents a 90\%
confidence level upper limit.  Taking the 90\% c.l.\ upper limit $R_c \le 1.20$,
one finds $\gamma \le 88^\circ$, which is consistent but not competitive
with other bounds on $\gamma$ \cite{CKMfitter}.

Writing a first order expression in $r_c$,
\beq\label{eqn:asymB+}
A_{CP}(B^+\to K^+\pi^0) \simeq -2r_c\sin\delta_c\sin\gamma~~,
\eeq
we can now compare this asymmetry with the asymmetry in $B^0\to K^+\pi^-$. 
The amplitude of this process~\cite{Gronau:1994rj,Gronau:1995hn},
\beq
-A(B^0\to  K^+\pi^-) = p + t~~,
\eeq
implies an asymmetry
\beq\label{eqn:asymB0}
A_{CP}(B^0\to K^+\pi^-) \simeq -2r\sin\delta\sin\gamma~~,
\eeq
where $r\equiv |T|/|p|$, and $\delta$ is the strong phase difference between 
$T$ and $p$. We have neglected an interference of an electroweak penguin 
amplitude with $T$ which is higher order and a term which is quadratic in $r$.

A comparison of Eqs.~(\ref{eqn:asymB+}) and (\ref{eqn:asymB0}) with current 
measurements of the two asymmetries can shed some light on $\delta_{CT}\equiv
{\rm arg}(C/T)$, the strong phase difference between the color-suppressed and
color-favored tree amplitudes.  An interesting question is whether this phase
difference can be small. A small phase, no larger than $15^\circ$, calculated in
QCD factorization for ${\rm arg}(C/T)$ in $B\to \pi\pi$~\cite{Beneke-private},
would be a useful input for extracting the weak
phase $\gamma$ in these decays~\cite{Bauer:2004tj}.  Assuming small SU(3)
breaking in the phase, one does not expect a much larger phase in $B\to K\pi$
than in $B\to \pi\pi$.  We note that in contrast to QCD factorization, large
negative values were obtained for the phase of $C/T$ in fits to $B\to K\pi$ and
$B\to \pi\pi$ data \cite{Chiang:2004nm,Buras:2003dj,Chiang:2006ih}. 

Let us assume for a moment that the magnitude of the phase difference
$\delta_{CT}$ is smaller than $\delta$, for which evidence of being nonzero and
positive has been obtained in $A_{CP}(B^0\to K^+\pi^-)$. [As an example,
$A_{CP}(B^0\to K^+\pi^-)=-0.10,~r=0.14$ and $\gamma=63^\circ$ correspond to
$\delta = 24^\circ$.] Simple geometry shows that $|T+C|\sin\delta_c  - |T|
\sin\delta = |C|\sin(\delta + \delta_{CT})$. That is, $r_c\sin \delta_c >
r\sin\delta$ holds for $\delta_{CT} > -\delta$. This implies 
that, if $\delta_{CT}$ is positive and of arbitrary size or negative but smaller
in magnitude than $\delta$, the asymmetry in $B^+\to K^+\pi^0$ should be of
the same sign (i.e. negative) and larger than the asymmetry in $B^0\to
K^+\pi^-$.  This is excluded within $4.7\sigma$ by the measured asymmetries. 

To conclude, we have proved a sum rule (\ref{eqn:sr}) which shows that one
can have both $R_c$ near 1 and $A_{CP}(B^+ \to K^+ \pi^0)$ near zero in the
presence of significant non-penguin amplitudes.  The key feature of the
sum rule is the approximate cancellation in $R_c$ between the real part of the
ratio $(T+C)/P$ and an electroweak penguin contribution.  We have used the
measured asymmetries in $B^+\to K^+\pi^0$ and $B^0\to K^+\pi^-$ to show that
the strong phase difference between the color-suppressed and color-favored tree
amplitudes contributing to $B^+\to K^+\pi^0$ 
must be negative and cannot be very small.
\medskip

We thank Martin Beneke, Harry  Lipkin and Dan Pirjol for helpful comments.
This work was supported in part by the Israel Science Foundation
under Grant No.\ 1052/04, by the German-Israeli Foundation under
Grant No.\ I-781-55.14/2003, and by the U. S. Department of Energy under
Grant No.\ DE-FG02-90ER40560.

\end{document}